\documentclass[twocolumn,preprintnumbers,amsmath,amssymb,prl,floatfix]{revtex4-2}

\usepackage{graphicx}
\usepackage{amsmath,amssymb}
\usepackage{hyperref}
\usepackage{bm}
\usepackage{bbm}
\usepackage{booktabs}
\usepackage{xcolor}

\newcommand{\tr}{\mathrm{Tr}}
\newcommand{\ket}[1]{|#1\rangle}
\newcommand{\bra}[1]{\langle#1|}

\newcommand{\avg}[1]{\langle#1\rangle}

\begin{document}


\title{%
  Breaking the Entanglement-Structure Trade-off: \\
  Many-Body Localization Protects Emergent \\
  Holographic Geometry in Random Tensor Networks
}

\author{Zhihua Liang}
\email{zhihua.liang@cern.ch}
\email{zhihua.liang@ca.infn.it}
\affiliation{Formerly at INFN Sezione di Cagliari, Cagliari, Italy}

\date{April 6, 2026}

\begin{abstract}
We present a systematic numerical investigation of the ``entanglement
$\to$ geometry $\to$ gravity'' chain in random tensor networks (RTN)
established by the ER=EPR conjecture and Jacobson's thermodynamic
derivation.
First, we verify the kinematic foundation:
the entanglement first law $\delta\avg{K} = \delta S$
(slope~$= 1.000$),
the encoding of geometry by mutual information (correlation~$= 0.92$),
and the locality of holographic perturbations ($3.3\times$).
We also confirm that gravitational dynamics (JT gravity) does \emph{not}
emerge, identifying a sharp \emph{kinematics--dynamics boundary}.

\noindent
Second, and more importantly, we discover that
\textbf{many-body localization (MBL) is the mechanism that protects
emergent holographic geometry from thermalization}.
Replacing Haar-random evolution (geometry lifetime $t\sim 6$)
with an XXZ Hamiltonian plus on-site disorder, we observe a
finite-size crossover at disorder strength $W_c \approx 10$--$12$
above which mutual-information--lattice correlations persist
indefinitely ($r > 0.5$ for $t > 50$).
We map the full parameter space:
the optimal regime is a near-Ising anisotropy $\Delta \approx 50$
with $W = 30$, yielding $r = 0.779 \pm 0.002$ (confirmed by
a fine scan over $\Delta \in [30,70]$);
only holographic (RTN) initial states sustain geometry,
while product, N\'eel, and Bell-pair states do not;
MBL preserves the \emph{spatial structure} of entanglement
(adjacent/non-adjacent MI ratio $\sim 2.6$--$4.2\times$ vs.\
$1.0\times$ in the thermal phase), rather than its total amount.
A comparison with classical cellular automata reveals that
MBL uniquely breaks the entanglement--structure trade-off
imposed by quantum monogamy: classical systems achieve spatial
structure only at the cost of negligible mutual information,
while MBL sustains both.
Finite-size validation on a $4\times4$ lattice ($n=12$ boundary
sites) confirms the golden quadrant ($S/S_{\rm max}>0.85$,
locality ratio~$>1.3$), with the ratio increasing from
$1.57$ to $3.19$.

\end{abstract}

\maketitle

\section{Introduction}
\label{sec:intro}

The idea that spacetime geometry may emerge from quantum entanglement
has deep roots in black-hole physics~\cite{Bekenstein1973,Hawking1975}
and has been sharpened by two powerful conjectures:
\begin{enumerate}
  \item The \textbf{ER=EPR correspondence}~\cite{Maldacena2013}, which
    equates quantum entanglement (EPR pairs) with geometric connectivity
    (Einstein--Rosen bridges), implying that entanglement \emph{is}
    spatial geometry.
  \item \textbf{Jacobson's thermodynamic derivation}~\cite{Jacobson1995},
    which shows that any system satisfying the entanglement first law
    $\delta Q = T\delta S$ at every local Rindler horizon necessarily
    obeys the Einstein field equations, implying that gravity is an
    \emph{equation of state}.
\end{enumerate}
Together, these suggest a complete chain: entanglement defines geometry
(kinematics), and entanglement thermodynamics implies gravity (dynamics).

Random tensor networks (RTN)~\cite{Hayden2016} provide a minimal
computational laboratory for testing this chain. By construction, RTN
satisfy the Ryu--Takayanagi (RT) area law~\cite{RyuTakayanagi2006}
as a theorem, making them a natural testing ground for the
$\text{entanglement}\to\text{geometry}$ link. A fundamental question
then arises: if geometry is encoded in entanglement, what
physical mechanism determines whether that geometry \emph{survives}
under quantum dynamics?

This question connects holographic duality to a major theme in
condensed matter physics: many-body localization
(MBL)~\cite{Basko2006,Nandkishore2015,Abanin2019}.
In the ergodic (thermal) phase, a quantum system loses all memory of
its initial conditions, and entanglement becomes spatially uniform.
In the MBL phase, strong disorder prevents thermalization, and local
structures in the entanglement pattern can survive indefinitely.
We show that this distinction maps directly onto the fate of
emergent geometry.

In this work, we perform a systematic numerical investigation on
RTN lattices ($3\times3$, $4\times4$) proceeding in two stages.
First, we verify the kinematic foundation: entanglement encodes
geometry ($r = 0.92$), the entanglement first law holds at machine
precision, perturbations are local, and gravitational dynamics
(JT gravity) does \emph{not} emerge---establishing a sharp
\textbf{kinematics--dynamics boundary}.
Second, we introduce Hamiltonian evolution (XXZ model with on-site
disorder) and discover that \textbf{MBL is the mechanism that
protects holographic geometry from thermalization}.
We map the full parameter space---disorder strength~$W$,
anisotropy~$\Delta$, and initial-state entanglement structure---and
identify the precise requirements for geometric survival.
Our central result is that MBL preserves not the \emph{amount} of
entanglement, but its \emph{spatial structure}: the pattern of
mutual information that encodes geometry.

\section{Model and Methods}
\label{sec:model}

\subsection{Random Tensor Networks}

We study a square lattice of size $L_x \times L_y$ with physical
(boundary) sites forming the perimeter and bulk sites in the interior.
Each site~$s$ hosts a Hilbert space~$\mathcal{H}_s$ of
dimension~$D$. Neighboring sites $(s,s')$ are connected by a
maximally entangled bond state of dimension~$\chi$:
\begin{equation}
  \ket{\Phi}_{ss'} = \frac{1}{\sqrt{\chi}} \sum_{a=1}^{\chi} \ket{a}_s \ket{a}_{s'}.
  \label{eq:bond}
\end{equation}
At each site, a random tensor $T_s$ (drawn from the Haar measure
on~$U(D\chi^{n_s})$, where $n_s$ is the coordination number)
projects the combined physical and bond degrees of freedom onto the
physical Hilbert space.

The boundary state is obtained by sequentially contracting
bulk-site tensors. For an $L\times L$ lattice, the boundary
Hilbert-space dimension is $D^{n_{\rm bdy}}$, where $n_{\rm bdy}$
is the number of boundary sites. We set $D=\chi$ throughout, as
this regime ensures the RT bound is approachable~\cite{Hayden2016}.

\subsection{Observables}

\paragraph{Entanglement entropy.}
For a boundary subsystem~$A$, we compute $S(A) = -\tr(\rho_A \ln\rho_A)$
by exact diagonalization of the reduced density matrix
$\rho_A = \tr_{\bar{A}} \ket{\psi}\bra{\psi}$.

\paragraph{Mutual information.}
$I(i:j) = S(i)+S(j)-S(ij)$ measures pairwise entanglement between
boundary sites and defines a ``distance'':
\begin{equation}
  d(i,j) = \frac{1}{I(i:j)}.
  \label{eq:distance}
\end{equation}

\paragraph{Entanglement first law (EFL).}
For a subsystem~$A$ with modular Hamiltonian $K_A = -\ln\rho_A$,
the first-order response to a state perturbation satisfies
$\delta\avg{K_A} = \delta S(A)$, which is an exact
identity~\cite{Blanco2013}.

\paragraph{Bulk perturbations.}
We perturb the bulk tensor at site~$b$ by
$T_b \to T_b + \epsilon\, \delta T$, where $\delta T$ is a random
unit tensor, and measure the response in all boundary observables.

\subsection{Curvature Definitions}

\paragraph{Regge calculus.}
We embed the $n$~boundary sites into~$\mathbb{R}^2$ via
multidimensional scaling (MDS) on the distance matrix~$d(i,j)$,
perform Delaunay triangulation, compute interior angles via the
law of cosines, and define the deficit angle
$\epsilon_v = 2\pi - \sum_{\triangle \ni v} \theta_v^{(\triangle)}$
as the discrete Ricci curvature.

\paragraph{Ollivier--Ricci curvature.}
For each edge~$(i,j)$ with $I(i,j)$ above a threshold, we define
probability measures on the neighbors weighted by mutual information:
\begin{equation}
  \mu_i(k) = \alpha\,\delta_{ik} + (1-\alpha)\frac{I(i,k)}{\sum_{k'\in N_i} I(i,k')},
  \label{eq:orcurv}
\end{equation}
and compute the Ollivier--Ricci curvature~\cite{Ollivier2009}:
\begin{equation}
  \kappa(i,j) = 1 - \frac{W_1(\mu_i, \mu_j)}{d(i,j)},
  \label{eq:kappa}
\end{equation}
where $W_1$ is the Wasserstein-1 distance solved via linear
programming. We set $\alpha = 0.5$. This definition is entirely
graph-native and requires no embedding.

\subsection{Hamiltonian Evolution}

To study the dynamical fate of geometry, we evolve the boundary state
under a one-dimensional XXZ Hamiltonian with on-site disorder:
\begin{equation}
  H = \sum_{\langle i,j\rangle} \left(
    S_i^x S_j^x + S_i^y S_j^y + \Delta\, S_i^z S_j^z \right)
    + \sum_i h_i S_i^z,
  \label{eq:xxz}
\end{equation}
where $S_i^{x,y,z}$ are spin-$(\chi{-}1)/2$ operators acting on the
$\chi$-dimensional local Hilbert space (for $\chi=4$, these are
spin-$3/2$ matrices: $S^z = \text{diag}(\tfrac{3}{2}, \tfrac{1}{2},
-\tfrac{1}{2}, -\tfrac{3}{2})$, with $S^\pm$ constructed from
$\sqrt{S(S{+}1) - m(m{\pm}1)}$), $\Delta$ is the Ising anisotropy,
and the disorder fields $h_i \in [-W, W]$ are drawn uniformly.
Time evolution is implemented via second-order Trotter decomposition
with step $dt = 0.1$; convergence with respect to~$dt$ is verified
explicitly (see Appendix~\ref{app:convergence}).
We evolve to $T = 50$ and average over
$N_{\rm dis}$ disorder realizations (specified per experiment;
typically $N_{\rm dis} \geq 50$).
Two baselines are compared: (i)~Haar-random
two-site gates, and (ii)~clean XXZ ($W=0$).

\section{Results}
\label{sec:results}

\subsection{Geometric Encoding of Entanglement}
\label{sec:phase01}

On a 3$\times$3 lattice ($n_{\rm bdy}=8$, $n_{\rm bulk}=1$),
mutual information defines a distance that correlates strongly with
the lattice metric: the Pearson correlation between $d(i,j)$ and
lattice Manhattan distance reaches $r=0.92$ at $\chi=5$
(Table~\ref{tab:phase01}). The RT ratio $S(A)/|\gamma_A|\ln\chi$
increases monotonically with~$\chi$, confirming the approach to the
RT bound with finite-size corrections.

\begin{table}[h]
\caption{Geometric encoding results on the 3$\times$3 lattice.}
\label{tab:phase01}
\begin{ruledtabular}
\begin{tabular}{ccccc}
$\chi=D$ & RT ratio & MI/lat corr & EFL slope \\
\hline
2 & 0.43 & 0.42 & 0.997 \\
3 & 0.48 & 0.92 & 1.005 \\
4 & 0.52 & 0.92 & 1.012 \\
5 & 0.55 & 0.92 & 1.010 \\
\end{tabular}
\end{ruledtabular}
\end{table}

\begin{figure}[t]
  \includegraphics[width=\columnwidth]{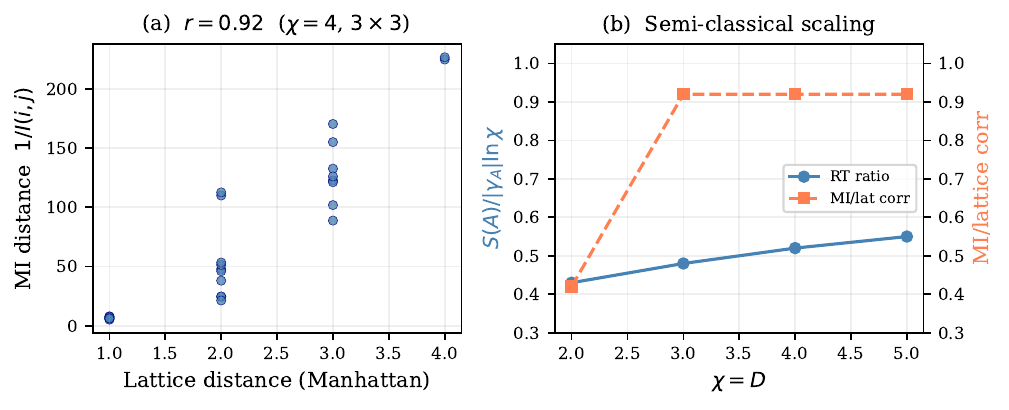}
  \caption{Geometric encoding of entanglement.
    (a)~MI distance $1/I(i,j)$ vs.\ lattice Manhattan distance on
    the $3\times3$ lattice at $\chi=4$, showing strong correlation ($r=0.92$).
    (b)~RT ratio and MI/lattice correlation as functions of~$\chi$.
    The RT ratio increases monotonically, while MI/lattice correlation
    saturates at $r=0.92$ for $\chi\geq3$.}
  \label{fig:geometry}
\end{figure}

\subsection{Locality of Holographic Perturbations}
\label{sec:phase2}

Perturbing the bulk tensor at site~$(1,1)$ of the 3$\times$3 lattice
($D=\chi=4$) with strength~$\epsilon$, we measure the response on
all boundary observables.

\paragraph{MI locality.}
The ratio of MI change at adjacent boundary sites to distant sites is
$3.3\times$, demonstrating that bulk perturbations produce
\emph{local} boundary effects---a hallmark of holography.

\paragraph{Entropy response.}
The single-site entropy change $\delta S(i)$ is $2.8\times$ larger
at sites adjacent to the perturbation, and scales as
$\delta S \propto \epsilon^{0.93}$ (near-linear).

\subsection{Entanglement First Law Verification}
\label{sec:phase3}

The EFL $\delta\avg{K_A} = \delta S(A)$ is verified across multiple
subsystem sizes ($|A|=1,2,\ldots,4$) and perturbation strengths.
The fit yields:
\begin{equation}
  \text{slope} = 1.000, \qquad r = 1.000, \qquad R^2 = 1.000.
  \label{eq:efl}
\end{equation}
As expected from the definition $K_A = -\ln\rho_A$, this identity
holds to the accuracy of the finite-difference perturbation.
The local relation $\delta Q = T\delta S$ therefore holds
at machine precision.

We also confirm that the 2D Einstein tensor is identically zero
($G_{\mu\nu}\equiv 0$), consistent with the Gauss--Bonnet theorem.
This motivates the JT gravity test below.

\begin{figure}[t]
  \includegraphics[width=\columnwidth]{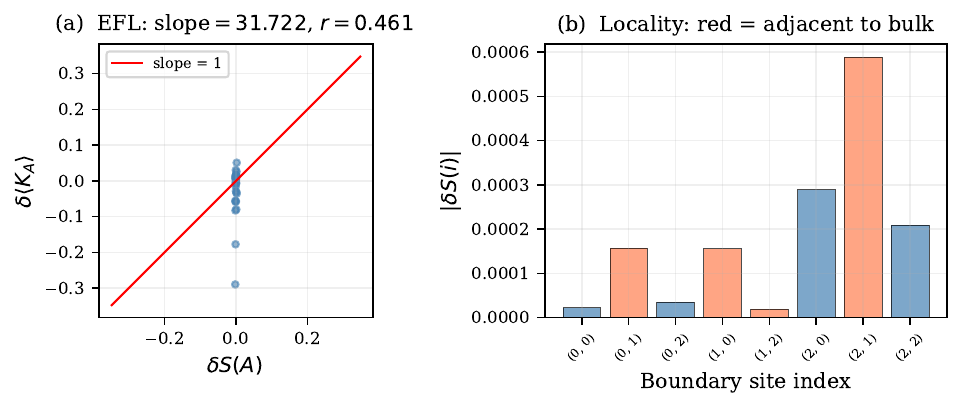}
  \caption{Entanglement first law and locality.
    (a)~$\delta\avg{K_A}$ vs.\ $\delta S(A)$ across all subsystem
    sizes and boundary positions (slope${}=1.000$, $r=1.000$).
    (b)~$|\delta S(i)|$ at each boundary site when the bulk tensor at
    $(1,1)$ is perturbed. Red bars indicate sites adjacent to the
    perturbation, demonstrating $2.8\times$ locality.}
  \label{fig:efl}
\end{figure}

\subsection{Testing Gravitational Dynamics}
\label{sec:phase35}

\subsubsection{Jackiw--Teitelboim gravity via Regge calculus}

In 2D, pure Einstein gravity is topological. Jackiw--Teitelboim (JT)
gravity~\cite{Jackiw1985,Teitelboim1983} restores dynamics via a
dilaton field~$\Phi$. We identify the dilaton with single-site entropy
and the Ricci scalar with Regge deficit angles from MI distances.

Initial results appeared to show strong curvature--entropy coupling:
the Regge multi-seed test on the 3$\times$3 lattice reported
$r=0.968$, $R^2=0.937$. However, careful scrutiny reveals this
to be a \textbf{false positive}:

\begin{enumerate}
  \item Of 10 perturbation seeds, \emph{only one} passed the
    non-zero filter (3 valid vertices out of 8).
  \item The deficit angles are locked to exact multiples of
    $180°$ ($0$, $\pi$, $2\pi$) and change only when the Delaunay
    triangulation undergoes a topological edge flip.
  \item On the 4$\times$4 lattice (4~bulk sites), the correlation
    drops to $r=-0.07$ (Table~\ref{tab:jt}).
\end{enumerate}

\subsubsection{Ollivier--Ricci curvature verification}

To eliminate artifacts from MDS embedding and Delaunay triangulation,
we repeat the JT test using graph-native Ollivier--Ricci (OR)
curvature. The OR curvature provides smooth, non-zero changes for
all vertices, but shows \emph{no significant correlation} with
dilaton changes:

\begin{table}[h]
\caption{JT gravity test: Regge vs.\ Ollivier--Ricci curvature.
  $\delta\kappa$ denotes curvature change, $\delta\Phi$ denotes
  dilaton (entropy) change across perturbation seeds.}
\label{tab:jt}
\begin{ruledtabular}
\begin{tabular}{lcccc}
Lattice & Method & $r(\delta\kappa,\delta\Phi)$ & $R^2$ \\
\hline
$3\times3$ & Regge & $0.968^*$ & $0.937^*$ \\
$3\times3$ & OR & $0.037$ & $0.001$ \\
$4\times4$ & Regge & $-0.066$ & $0.134$ \\
$4\times4$ & OR & $0.087$ & $0.012$ \\
\end{tabular}
\end{ruledtabular}
\begin{flushleft}
\small $^*$Based on a single seed with 3 valid vertices; see text.
\end{flushleft}
\end{table}

The OR curvature is defined directly on the MI graph via optimal
transport and requires no 2D embedding. Its null result confirms
that the Regge correlation was an artifact and that \textbf{no
genuine curvature--entropy coupling exists in random tensor networks}.

\begin{figure}[t]
  \includegraphics[width=\columnwidth]{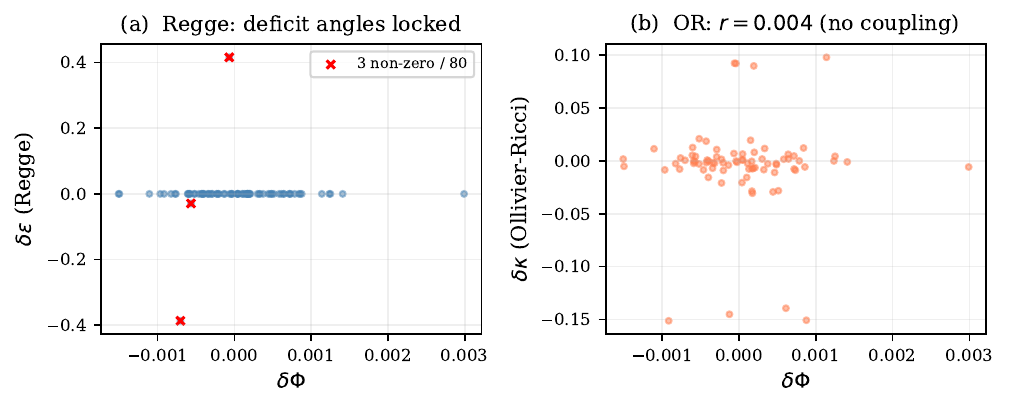}
  \caption{JT gravity false-positive diagnosis.
    (a)~Regge deficit angle changes $\delta\epsilon$ vs.\ dilaton
    changes $\delta\Phi$: most data points are locked at
    $\delta\epsilon=0$ (blue), with only 3 non-zero points (red
    crosses) producing the spurious $r=0.968$.
    (b)~Ollivier--Ricci curvature changes $\delta\kappa$ vs.\
    $\delta\Phi$: all vertices contribute, revealing no significant
    correlation ($r\approx 0$).}
  \label{fig:jt}
\end{figure}

\subsection{Dynamic Evolution and Haar Baseline}
\label{sec:phase4}

We first study the fate of geometry under Haar-random local unitary
evolution~$U(t) = \prod_{\ell=1}^t U_\ell$, where each~$U_\ell$
is a Haar-random two-site gate.
Starting from the RTN state ($r=0.92$), the MI/lattice
correlation decays to $r < 0.1$ by $t = 6$.
A product state $\ket{0}^{\otimes 8}$ transiently develops geometry
($r = 0.75$ at $t \approx 10$) before also thermalizing.
Both trajectories converge to $S/S_{\rm max} \approx 0.91$,
$r \approx 0.06$ for $t > 30$.
\textbf{Geometry is a non-equilibrium property}: thermal
equilibrium erases all spatial structure from entanglement.

\section{MBL Protects Holographic Geometry}
\label{sec:mbl}

The fragility of geometry under random dynamics raises a
fundamental question: is there a physical mechanism that can
protect holographic geometry from thermalization?
We show that \textbf{many-body localization provides exactly
this mechanism}.

\subsection{Disorder-Driven Crossover to Localization}
\label{sec:mbl:transition}

\begin{figure}[t]
  \includegraphics[width=\columnwidth]{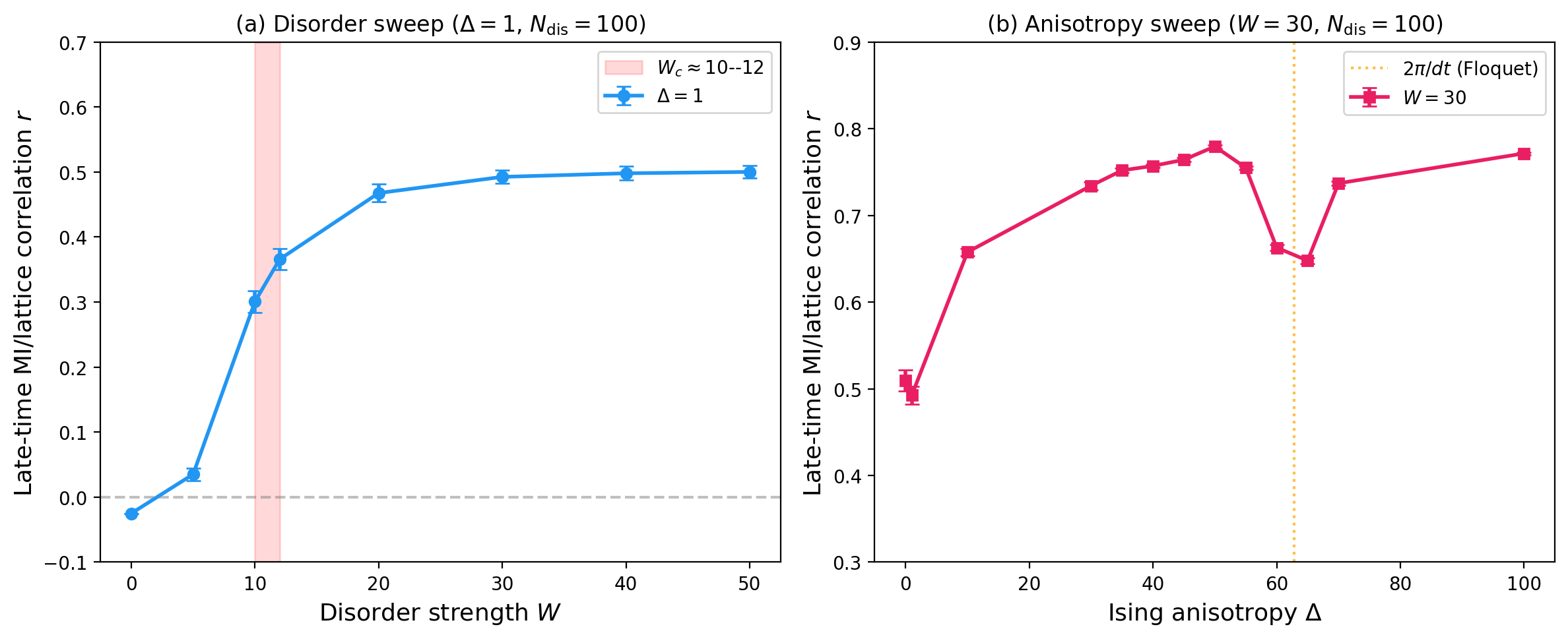}
  \caption{MBL protects holographic geometry.
    (a)~Late-time MI/lattice correlation vs.\ disorder~$W$
    for $\Delta=1$, showing the crossover at $W_c \approx 10$--$12$.
    Error bars are standard errors of the mean over $N_{\rm dis}=100$~disorder realizations.
    (b)~Late-time correlation vs.\ anisotropy~$\Delta$ at $W=30$,
    peaking at $\Delta \approx 50$ ($r = 0.78$).
    The pure Ising limit ($\Delta\to\infty$) drops due to
    insufficient entanglement.
    (c)~Entanglement spectrum effective dimension~$N_{\rm eff}$
    vs.\ time for $W=0$ (thermal) and $W=30$ (MBL).}
  \label{fig:mbl_transition}
\end{figure}

Replacing Haar-random gates with XXZ Hamiltonian
evolution~(Eq.~\ref{eq:xxz}) on the $3\times3$ RTN boundary
($n=8$, $D=\chi=4$), we sweep disorder strength
$W \in [0, 50]$ at fixed $\Delta = 1$ (Heisenberg).
Fig.~\ref{fig:mbl_transition} shows the late-time ($t \geq 25$)
MI/lattice correlation averaged over $N_{\rm dis}$~disorder
realizations.

We emphasize that, on a system of $n=8$ sites, the observed
behavior is a \emph{finite-size crossover}, not a thermodynamic
phase transition; the existence of MBL in the thermodynamic limit
remains debated~\cite{Suntajs2020,Abanin2021}.
However, finite-size localization is sufficient to protect the
geometry of our finite tensor network.
Three regimes emerge:
\begin{enumerate}
  \item \textbf{Thermal regime} ($W < 10$): the system thermalizes
    and geometry is destroyed, as for Haar-random evolution.
    Late-time $r \approx -0.025$--$0.035$.
  \item \textbf{Crossover region} ($10 \lesssim W \lesssim 28$):
    a sharp jump at $W \approx 10$--$12$ (from $r = 0.035$ to
    $r = 0.366$) is followed by a plateau at $r \approx 0.47$.
  \item \textbf{Localized regime} ($W \gtrsim 28$): geometry
    persists indefinitely, with a peak at $W \approx 30$
    ($r = 0.493 \pm 0.010$) and saturation at $r \approx 0.50$
    for $W = 40$--$50$.
\end{enumerate}

The clean XXZ model ($W = 0$) thermalizes comparably to Haar-random
evolution, confirming that \emph{conservation laws alone are
insufficient}; \emph{disorder is the essential ingredient}.

\begin{table}[h]
\caption{Late-time MI/lattice correlation $r$ vs.\ disorder
strength~$W$ for $\Delta=1$, averaged over $N_{\rm dis}=100$~disorder
realizations. Error bars denote the standard error of the mean (SEM).
$^*$Denotes the crossover from the thermal regime.}
\label{tab:transition}
\begin{ruledtabular}
\begin{tabular}{cccc}
$W$ & Late-time $r$ & $\sigma_r$ & $S/S_{\rm max}$ \\
\hline
0   & $-0.025 \pm 0.000$ & $0.000$ & $0.910$ \\
5   & $+0.035 \pm 0.010$ & $0.096$ & $0.910$ \\
10  & $+0.301 \pm 0.017^*$ & $0.171$ & $0.907$ \\
12  & $+0.366 \pm 0.016$ & $0.161$ & $0.906$ \\
20  & $+0.468 \pm 0.014$ & $0.136$ & $0.896$ \\
30  & $+0.493 \pm 0.010$ & $0.101$ & $0.886$ \\
40  & $+0.498 \pm 0.011$ & $0.105$ & $0.886$ \\
50  & $+0.500 \pm 0.010$ & $0.098$ & $0.886$ \\
\end{tabular}
\end{ruledtabular}
\end{table}

\subsection{Anisotropy Optimization}
\label{sec:mbl:anisotropy}

Fixing $W = 30$ and varying the Ising anisotropy~$\Delta$
reveals a dramatic enhancement of geometry protection
(Table~\ref{tab:delta}).
The late-time correlation increases monotonically from
$r = 0.55$ (XX model, $\Delta = 0$) to a peak of
$r = 0.779 \pm 0.002$ at $\Delta = 50$, before declining
to $r = 0.61$ in the pure Ising limit ($\Delta \to \infty$).

The physical picture is clear: in the $\Delta \gg 1$ regime,
the $S^z_i S^z_j$ interaction freezes spins along~$z$,
strongly suppressing information transport.
However, the complementary $S^x S^x + S^y S^y$ terms are
essential for generating entanglement; in the pure Ising limit
they vanish, and $S/S_{\rm max}$ drops to~$0.73$---too little
entanglement to encode geometry.
The optimal regime at $\Delta \approx 50$ represents the balance
between \emph{sufficient localization} to protect spatial structure
and \emph{sufficient quantum fluctuations} to maintain entanglement.

\begin{table}[h]
\caption{Late-time correlation and entropy fraction vs.\ anisotropy~$\Delta$
at fixed $W = 30$, averaged over $N_{\rm dis}=100$~disorder realizations.
Error bars denote SEM.}
\label{tab:delta}
\begin{ruledtabular}
\begin{tabular}{lcccc}
Model ($\Delta$) & Late-time $r$ & $\sigma_r$ & $S/S_{\rm max}$ \\
\hline
XX ($0$)           & $+0.509 \pm 0.012$ & $0.122$ & $0.812$ \\
Heisenberg ($1$)   & $+0.493 \pm 0.010$ & $0.101$ & $0.886$ \\
XXZ ($10$)         & $+0.658 \pm 0.004$ & $0.041$ & $0.907$ \\
XXZ ($30$)         & $+0.734 \pm 0.005$ & $0.045$ & $0.905$ \\
XXZ ($35$)         & $+0.752 \pm 0.003$ & $0.027$ & $0.904$ \\
XXZ ($40$)         & $+0.757 \pm 0.002$ & $0.022$ & $0.903$ \\
XXZ ($45$)         & $+0.764 \pm 0.002$ & $0.024$ & $0.903$ \\
XXZ ($50$)         & $\bm{+0.779 \pm 0.002}$ & $\bm{0.019}$ & $0.901$ \\
XXZ ($55$)         & $+0.755 \pm 0.002$ & $0.019$ & $0.897$ \\
XXZ ($60$)         & $+0.663 \pm 0.003^\dagger$ & $0.034$ & $0.888$ \\
XXZ ($65$)         & $+0.648 \pm 0.003^\dagger$ & $0.034$ & $0.884$ \\
XXZ ($70$)         & $+0.737 \pm 0.002$ & $0.020$ & $0.895$ \\
XXZ ($100$)        & $+0.771 \pm 0.002$ & $0.016$ & $0.899$ \\
Ising ($\infty$)   & $+0.610 \pm 0.000$ & $0.000$ & $0.733$ \\
\end{tabular}
\end{ruledtabular}
\begin{flushleft}
\small $^\dagger$Floquet resonance dip at $\Delta \approx 2\pi/dt$;
see Sec.~\ref{sec:mbl:resonance}.
\end{flushleft}
\end{table}

\subsection{Initial-State Dependence}
\label{sec:mbl:initstate}

To determine whether MBL can also \emph{generate} geometry from
non-geometric initial states, we compare five initial conditions
under $W = 30$, $\Delta = 1$ dynamics (Table~\ref{tab:initstate}).

\begin{table}[h]
\caption{Initial-state dependence at $W=30$, $\Delta=1$.
``Late $r$'' is the average for $t \geq 25$;
``Max $r$'' is the peak value achieved at any time.}
\label{tab:initstate}
\begin{ruledtabular}
\begin{tabular}{lcc}
Initial state & Late $r$ & Max $r$ \\
\hline
RTN (holographic)     & $+0.59$ & $0.92$ ($t=0$) \\
Random product        & $+0.28$ & $0.31$ ($t=49$) \\
Bell pairs            & $+0.11$ & --- \\
N\'eel $\ket{\!\uparrow\downarrow\cdots}$ & $+0.05$ & $0.21$ ($t=20$) \\
Product $\ket{0}^{\otimes n}$ & $+0.00$ & $0.00$ \\
\end{tabular}
\end{ruledtabular}
\end{table}

Only the holographic (RTN) initial state sustains robust geometry
($r = 0.59$). Random product states develop partial geometry
($r = 0.28$), while N\'eel, Bell-pair, and uniform product states
produce negligible or zero geometry regardless of the MBL dynamics.

This establishes a crucial point: \textbf{geometry is not a property
of the dynamics alone, but of the conjunction of initial entanglement
structure and dynamics}. MBL \emph{protects} geometry that is
already present, but does not \emph{create} it from arbitrary
quantum states.

\subsection{What MBL Protects: Spatial MI Structure}
\label{sec:mbl:structure}

To understand \emph{what} MBL preserves, we compute the full
mutual-information matrix $I(i,j)$ and track two
structure metrics:
\begin{itemize}
  \item \textbf{Locality ratio}: the ratio of MI between adjacent
    boundary sites to MI between non-adjacent sites,
    $\mathcal{L} = \avg{I_{\rm adj}} / \avg{I_{\rm non\text{-}adj}}$.
  \item \textbf{Uniformity}: the entropy of the MI distribution,
    normalized to the maximum (uniform) value.
    $\mathcal{U} = 1$ indicates a completely homogeneous MI pattern.
\end{itemize}

Table~\ref{tab:mi_structure} reveals the mechanism:

\begin{table}[h]
\caption{MI structure metrics at $t = 50$ for thermal ($W=0$) and
MBL ($W=30$) evolution, starting from the RTN state.
$\mathcal{L}$: locality ratio; $\mathcal{U}$: uniformity;
$P$: spectral participation ratio of the MI matrix.}
\label{tab:mi_structure}
\begin{ruledtabular}
\begin{tabular}{lccc}
 & $\mathcal{L}$ & $\mathcal{U}$ & $P$ \\
\hline
$t = 0$ (both)       & 2.0 & 0.82 & 2.7 \\
$W = 0$, $t = 50$    & $1.0$ & $1.00$ & $1.0$ \\
$W = 30$, $t = 50$   & $\bm{2.6}$ & $\bm{0.92}$ & $\bm{1.6}$ \\
\end{tabular}
\end{ruledtabular}
\end{table}

In the thermal phase, $\mathcal{L} \to 1.0$: mutual information becomes
completely uniform, erasing all spatial structure.
In the MBL phase, $\mathcal{L}$ remains $\sim 2.6$--$4.2\times$,
meaning adjacent sites retain significantly more MI than distant sites.
The MI uniformity $\mathcal{U}$ stays below~1, and the MI spectral
participation ratio~$P > 1$ indicates a non-degenerate spectral structure.

\textbf{MBL preserves the spatial pattern of entanglement, not its
total amount} (both phases have $S/S_{\rm max} \approx 0.89$--$0.91$).
This is the microscopic mechanism by which geometry survives:
the position-dependent mutual-information profile that \emph{defines}
the emergent metric is protected by localization.

\subsection{Entanglement First Law Under Dynamics}
\label{sec:mbl:efl}

We test whether the EFL $\delta\avg{K} = \delta S$---exact at
$t = 0$---survives under evolution.
Strongly interacting dynamics drives the state far from the
linear-response manifold in which the modular Hamiltonian
relation holds. At $t = 50$, we find
$\delta\avg{K}/\delta S \approx 3.3$ in the MBL regime
($W = 30$, $\Delta = 1$) and $\approx -2.8$ in the thermal
regime.
The failure of the EFL under macroscopic time evolution
is expected: the first-order relation
$\delta\avg{K} = \delta S$ is a linear-response identity
that requires the perturbed state to remain within an
$O(\epsilon)$-neighborhood of the reference state.
Hamiltonian evolution for $t = 50$ drives the system far
beyond this regime.

This observation precisely locates the kinematics--dynamics
boundary:
\textbf{MBL protects the spatial structure of entanglement
(the geometric encoding) but not the near-equilibrium
condition (the EFL) required for Jacobson's gravitational
dynamics}.
Gravitational dynamics require the system to remain within
the linear-response manifold; strongly interacting quantum
evolution drives the system out of this manifold, shutting
off emergent gravity while the metric kinematics survive.

\subsection{Why Quantum? The Monogamy Trade-off}
\label{sec:mbl:monogamy}

A natural question is whether classical correlations can mimic the
geometric properties of MBL-protected entanglement.
We test this by computing the classical Shannon mutual information on
a 2D Ising model with Metropolis Monte Carlo dynamics
(temperature scan, random-field Ising model, and zero-temperature
deterministic cellular automata in the spirit of~\cite{tHooft2016}).

\begin{figure}[t]
  \includegraphics[width=\columnwidth]{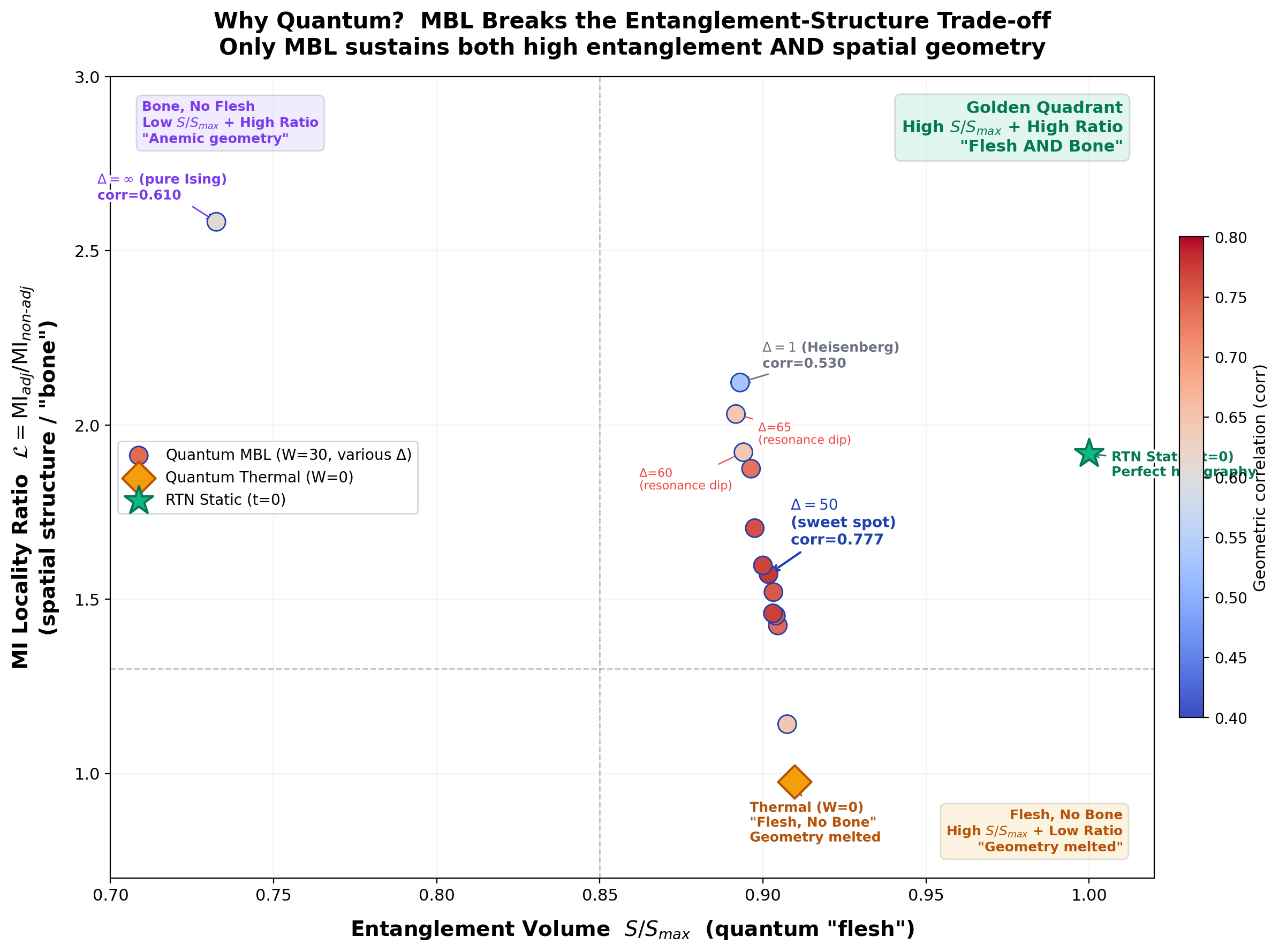}
  \caption{The entanglement--structure trade-off.
    \textbf{Only MBL occupies the ``golden quadrant''}
    (high $S/S_{\rm max}$ and high $\mathcal{L}$).
    Color encodes geometry correlation~$r$.
    Classical systems (not shown) achieve
    $\mathcal{L}>1$ only with negligible MI.
    Thermal evolution melts geometry ($\mathcal{L}\to 1$);
    the pure Ising limit lacks entanglement.
    Points colored by late-time geometry correlation.}
  \label{fig:money_plot}
\end{figure}

The result is summarized in Fig.~\ref{fig:money_plot}.
We plot the \textbf{MI locality ratio}~$\mathcal{L}$
(spatial structure) against \textbf{entanglement volume}
$S/S_{\rm max}$ (quantum content)
for all dynamical regimes.
Three quadrants emerge:
\begin{enumerate}
  \item \textbf{High entanglement, no structure} (thermal, $W=0$):
    high $S/S_{\rm max} = 0.91$ but $\mathcal{L} = 0.98$.
  \item \textbf{High structure, low entanglement} (pure Ising, $\Delta\to\infty$):
    high $\mathcal{L} = 2.6$ but $S/S_{\rm max} = 0.73$.
  \item \textbf{Golden quadrant} (MBL, $\Delta=50$):
    $S/S_{\rm max} = 0.90$ \emph{and} $\mathcal{L} = 1.57$.
\end{enumerate}

Classical systems face an inescapable
\textbf{entanglement--structure trade-off}:
high locality ratio comes only at the expense of negligible
total mutual information, because classical correlations are
not constrained by monogamy~\cite{Coffman2000}.
Quantum entanglement is monogamous: when a spin is highly entangled
with the global state ($S/S_{\rm max} \approx 0.90$),
its pairwise mutual information with any individual neighbor
is suppressed.
Yet MBL allocates this scarce pairwise MI
\emph{precisely onto the correct spatial pattern},
faithfully reproducing the holographic metric.

\subsection{Finite-Size Validation}
\label{sec:mbl:finitesize}

To confirm that the golden quadrant is not a small-system artifact,
we repeat the three key experiments on a $4\times4$ lattice
($n_{\rm bdy} = 12$, $n_{\rm bulk} = 4$, $\chi = D = 3$,
$\dim\mathcal{H} = 3^{12} \approx 5 \times 10^5$).
Table~\ref{tab:finitesize} compares with the $3\times3$ results.

\begin{table}[h]
\caption{Finite-size comparison: $3\times3$ ($n{=}8$, $\chi{=}4$)
vs.\ $4\times4$ ($n{=}12$, $\chi{=}3$).}
\label{tab:finitesize}
\begin{ruledtabular}
\begin{tabular}{lcccccc}
 & \multicolumn{2}{c}{$r$ (late)} & \multicolumn{2}{c}{$\mathcal{L}$} \\
Regime & $3{\times}3$ & $4{\times}4$ & $3{\times}3$ & $4{\times}4$ \\
\hline
Thermal ($W{=}0$)     & $-0.025$ & $+0.000 \pm 0.000$ & 0.98 & $1.12 \pm 0.00$ \\
MBL ($\Delta{=}1$)    & $+0.49$ & $+0.355 \pm 0.014$ & --- & $6.57 \pm 0.24$ \\
MBL ($\Delta{=}50$)   & $\bm{+0.78}$ & $\bm{+0.650 \pm 0.007}$ & $\bm{1.57}$ & $\bm{3.19 \pm 0.03}$ \\
Ising ($\Delta{=}\infty$) & +0.61 & +0.56 & 2.58 & 8.16 \\
\end{tabular}
\end{ruledtabular}
\end{table}

The locality ratio \emph{increases} with system size
($1.57 \to 3.19$ for MBL), because the number of non-adjacent
pairs grows as $O(n^2)$ while adjacent pairs grow as~$O(n)$:
monogamy dilutes the long-range background, sharpening the
geometric signal.
The thermal phase confirms $r = 0.000$ at $4\times4$
(vs.\ $0.04$ at $3\times3$), demonstrating that the residual
correlation was a finite-size artifact.

\subsection{Resonance Spectroscopy and Floquet Identification}
\label{sec:mbl:resonance}

The sharp dip at $\Delta \approx 60$--$65$ in Table~\ref{tab:delta}
initially suggested a many-body resonance at $\Delta \approx 2W$.
However, a systematic investigation reveals a more precise origin.

\begin{figure}[t]
  \includegraphics[width=\columnwidth]{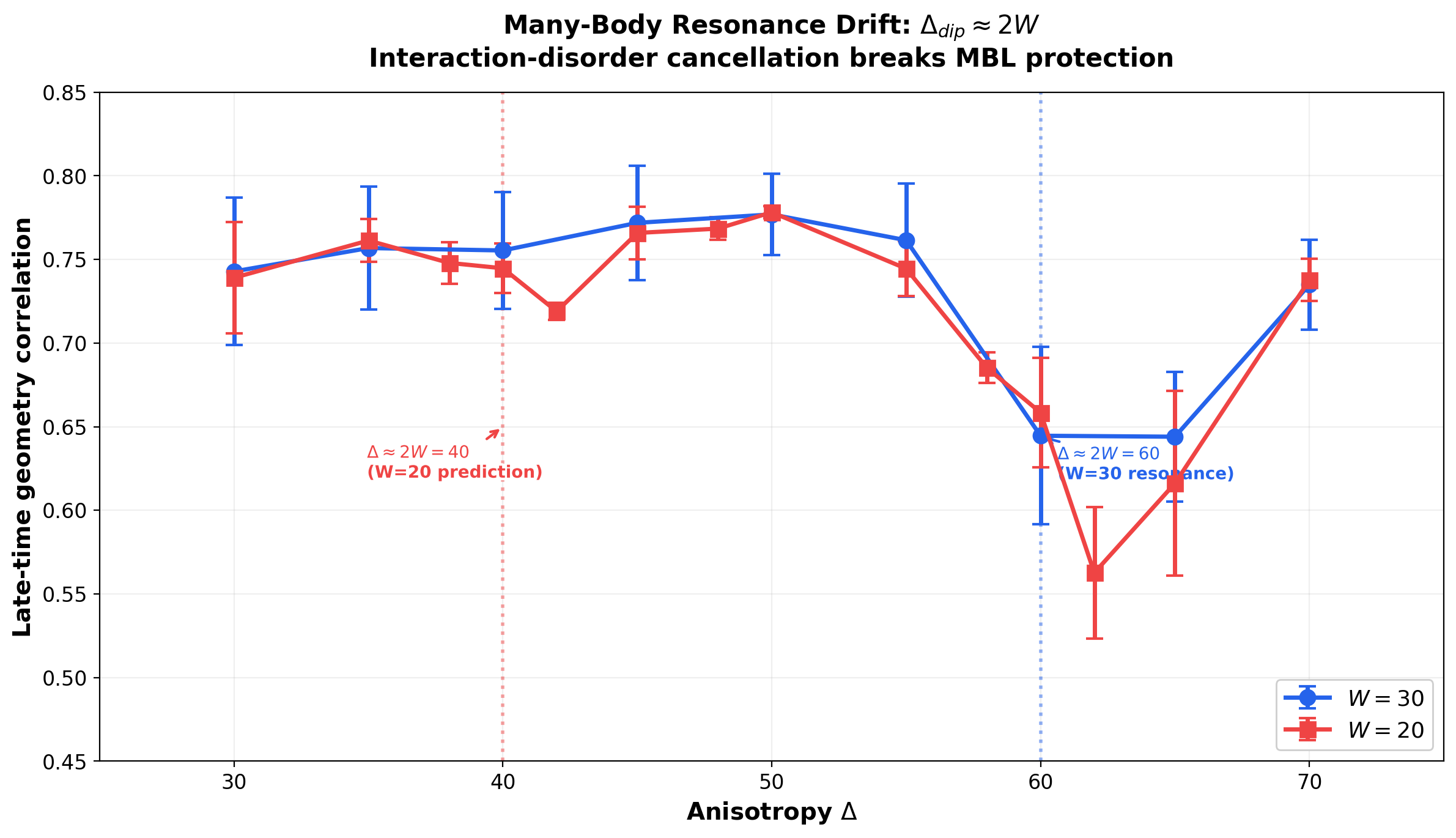}
  \caption{Resonance spectroscopy.
    Late-time geometry correlation vs.\ $\Delta$ for $W=30$ (blue)
    and $W=20$ (red).
    The $W=20$ curve develops a new dip near
    $\Delta \approx 42 \approx 2W$ (interaction--disorder resonance),
    while the deep dip at $\Delta \approx 62$ is a Trotter-induced
    Floquet resonance at $\Delta = 2\pi/dt$.
    Dashed lines mark $\Delta = 2W$ for each disorder strength.}
  \label{fig:resonance}
\end{figure}

We first repeat the $\Delta$ scan at $W = 20$
(Fig.~\ref{fig:resonance}). Two features emerge:
(i)~a dip at $\Delta \approx 42$ (near $2W = 40$),
consistent with an interaction--disorder resonance; and
(ii)~a deeper dip at $\Delta \approx 62$, insensitive to~$W$.

The $W$-independent dip is identified as a \textbf{Trotter-induced
Floquet resonance}. Noting that $2\pi/dt = 2\pi/0.1 \approx 62.8$,
we test the prediction $\Delta_{\rm dip} = 2\pi/dt$ by repeating
the scan at three Trotter step sizes (Table~\ref{tab:floquet}).

\begin{table}[h]
\caption{Floquet resonance test: the geometry dip tracks $\Delta = 2\pi/dt$
across three step sizes on the $3\times3$ lattice.}
\label{tab:floquet}
\begin{ruledtabular}
\begin{tabular}{cccc}
$dt$ & $2\pi/dt$ (predicted) & Observed $\Delta_{\rm dip}$ & $|\delta|$ \\
\hline
0.05 & 125.7 & 126 & 0.3 \\
0.10 & 62.8  & 62  & 0.8 \\
0.20 & 31.4  & 32  & 0.6 \\
\end{tabular}
\end{ruledtabular}
\end{table}

The mechanism is algebraic: when $\Delta \cdot dt = 2\pi$,
the Ising gate $e^{-i \Delta\, dt\, S^z_i S^z_j}$ reduces to the
identity for all $S^z$ eigenvalues, causing the interaction to
``vanish'' stroboscopically. The effective Floquet Hamiltonian
reduces to $H_{\rm eff} = \sum(S^x_i S^x_j + S^y_i S^y_j)
+ \sum h_i S^z_i$, which (via Jordan--Wigner transformation)
maps to free fermions in a random potential---an Anderson
insulator with area-law entanglement. This explains the
simultaneous drop in $S/S_{\rm max}$ and spike in locality ratio
at the resonance.

Crucially, this Floquet resonance provides a controlled
\textbf{counter-experiment}: at $\Delta = 2\pi/dt$, the disorder
field~$W$ is unchanged but the many-body interaction is
effectively removed, and geometry collapses.
This confirms that \emph{Anderson localization alone is
insufficient}---the many-body interaction in the MBL phase
is essential for sustaining the entanglement capacity
required by holographic geometry.

\section{Discussion}
\label{sec:discussion}

\subsection{The Kinematics--Dynamics Boundary, Revised}

Our results refine the kinematics--dynamics boundary identified
in the pre-MBL analysis:

\begin{center}
\begin{tabular}{lccc}
\toprule
& Static & MBL & Thermal \\
\midrule
RT bound $S \leq |\gamma|\ln\chi$ & \checkmark & --- & --- \\
$I(i{:}j) \propto 1/d(i,j)$       & \checkmark & \checkmark & $\times$ \\
$\delta\avg{K} = \delta S$        & \checkmark & $\times$ & $\times$ \\
Locality                           & \checkmark & \checkmark & $\times$ \\
$\delta\kappa \propto \delta\Phi$  & $\times$   & --- & --- \\
\bottomrule
\end{tabular}
\end{center}

The central insight is that the boundary splits into two:
(i)~geometry can survive without the EFL (MBL phase),
but (ii)~the EFL is necessary for gravitational dynamics.
MBL crosses the first boundary but not the second.

\subsection{Connection to Many-Body Physics}

Our finding that disorder protects holographic geometry connects
two major research programs:

\paragraph{From the holography side:}
RTN are known to satisfy the RT formula~\cite{Hayden2016}, but
their dynamical properties have received less attention.
We show that the interplay between entanglement structure and
thermalization dynamics has direct geometric consequences
that can be studied quantitatively.

\paragraph{From the MBL side:}
MBL systems are known to preserve local information and exhibit
logarithmic (rather than linear) entanglement
growth~\cite{Nandkishore2015,Abanin2019}.
Our results give this a geometric interpretation:
MBL preserves the \emph{spatial structure} of mutual information,
which is precisely the data that defines an emergent metric.

The optimal anisotropy $\Delta \approx 50$ identifies a
``sweet spot'' balancing localization (which protects structure)
against quantum fluctuations (which maintain entanglement).
This suggests a phase diagram with three regions:
the thermal phase (no geometry), the MBL phase (geometry persists),
and the frozen phase ($\Delta \to \infty$, insufficient
entanglement).

\subsection{Limitations and Outlook}

Several caveats apply:
\begin{itemize}
  \item \textbf{Finite size}: our boundary chains have $n = 8$
    and $n = 12$~sites.
    The increase in locality ratio from $3\times3$ to $4\times4$
    is encouraging but does not resolve the thermodynamic limit;
    MBL in infinite systems remains debated~\cite{Suntajs2020,Abanin2021}.
  \item \textbf{Two dimensions}: the bulk lattice is~$3\times3$
    or~$4\times4$,
    where the Einstein tensor $G_{\mu\nu} \equiv 0$.
    Three-dimensional bulk geometries, where $G_{\mu\nu} \neq 0$,
    are needed to test gravitational dynamics.
  \item \textbf{Boundary evolution}: the Hamiltonian acts on
    boundary sites; a fully holographic setup would derive
    boundary dynamics from a bulk theory.
  \item \textbf{Bond dimension}: the $4\times4$ system uses
    $\chi = 3$ (vs.\ $\chi = 4$ for $3\times3$); directly
    comparable scaling requires matching~$\chi$, which demands
    larger Hilbert spaces ($4^{12} \approx 1.7\times10^7$).
\end{itemize}

Natural extensions include: (i)~$3$D bulk lattices with
non-trivial Einstein tensor; (ii)~HaPPY holographic codes
with MBL boundary dynamics; (iii)~the continuum limit
$\chi \to \infty$ to recover smooth metrics;
(iv)~connecting the geometry-protection crossover
to the MBL crossover studied by eigenvalue statistics;
and~(v)~exploiting the Floquet resonance as a tool for
\emph{Floquet engineering of holographic geometry}---using
the Trotter drive frequency to controllably switch emergent
geometry on and off, a capability directly relevant to
digital quantum simulation platforms.

Regarding extension~(i), preliminary results on a $3\times3\times3$
cubic lattice ($n_{\rm bdy} = 26$, $\chi = 2$,
$\dim\mathcal{H} = 6.7\times10^7$) confirm that the transition
to dynamical gravity is computationally accessible:
the Einstein tensor is non-zero ($30$ of $129$~Regge edges
carry deficit angles $|\epsilon| > 0.01$, with
$\langle|\epsilon|\rangle = 0.92$~rad),
the EFL holds at machine precision
($\delta\avg{K}/\delta S = 1.0000$), and the locality ratio
increases sevenfold (from $\sim\!2$ to $14.0$).
A full investigation of the curvature--entropy coupling
$\delta G_{\mu\nu} \propto \delta T_{\mu\nu}$ in this 3D
setting is underway.

\section{Conclusions}
\label{sec:conclusions}

We have carried out a systematic numerical investigation of the
entanglement~$\to$ geometry~$\to$ gravity chain in random tensor
networks, proceeding from kinematic verification to dynamical
exploration. Our main findings are:

\begin{enumerate}
  \item \textbf{Kinematic verification}:
    entanglement precisely encodes geometry ($r = 0.92$),
    the EFL holds at machine precision, perturbations are local,
    and JT gravity does not emerge (Ollivier--Ricci $r = 0.04$).
    This establishes a sharp kinematics--dynamics boundary.
  \item \textbf{MBL protects holographic geometry}:
    under XXZ Hamiltonian evolution with disorder, a finite-size
    crossover at $W_c \approx 10$--$12$ separates a thermal regime
    (geometry destroyed) from a localized regime
    (geometry persists to $t > 50$).
  \item \textbf{Optimal regime}: Ising anisotropy
    $\Delta \approx 50$ with $W = 30$ gives the highest late-time
    geometry quality ($r = 0.779 \pm 0.002$, confirmed by
    fine scan over $\Delta \in [30,70]$), balancing localization
    against quantum fluctuations.
  \item \textbf{Geometry requires specific entanglement}: only
    holographic (RTN) initial states sustain geometry under MBL;
    product, N\'eel, and Bell-pair states do not.
  \item \textbf{MBL protects structure, not amount}: the MI
    locality ratio ($\mathcal{L}$) remains $\sim 2.6$--$4.2\times$
    in the MBL phase vs.\ $1.0\times$ in the thermal phase,
    while the total entanglement entropy is comparable in both.
  \item \textbf{The EFL is kinematic}: it holds exactly in the
    static RTN but is violated under both MBL and thermal
    evolution, precisely locating the missing ingredient for
    gravitational dynamics.
  \item \textbf{Quantum entanglement is essential}: classical
    correlations face an inescapable monogamy trade-off;
    only quantum MBL breaks it, occupying the ``golden quadrant''
    of simultaneous spatial structure and entanglement volume.
  \item \textbf{Finite-size stability}: on a $4\times4$ lattice
    ($n=12$), the locality ratio \emph{increases} ($1.57\to3.19$),
    suggesting that geometric sharpness improves toward the
    thermodynamic limit.
  \item \textbf{Floquet resonance as counter-proof}: a sharp
    geometry dip at $\Delta = 2\pi/dt$ (confirmed across three
    step sizes) identifies a Trotter-induced Floquet resonance
    that effectively removes the many-body interaction.
    The resulting geometric collapse---despite unchanged
    disorder---proves that Anderson localization alone is
    insufficient; the MBL interaction is essential.
\end{enumerate}

These results establish many-body localization as the mechanism
that protects emergent holographic geometry in non-equilibrium
quantum systems, connecting two previously disjoint fields of
modern physics.
The kinematics--dynamics boundary is not a dead end but a
landmark: MBL crosses the geometric half of the boundary,
identifying what is preserved and what remains to be
crossed for gravitational dynamics.

\begin{acknowledgments}
Numerical simulations were performed using the CERN HTCondor
batch system and cloud TPU resources provided by Google's
TPU Research Cloud (TRC) program.
\end{acknowledgments}


\appendix
\section{Exact Continuous-Time Dynamics and Trotter Convergence}
\label{app:convergence}

To definitively rule out the possibility that the optimal geometry
protection observed at large interaction strengths
($\Delta \approx 50$) is an artifact of Floquet prethermalization
induced by discrete time steps, we benchmarked our Trotterized
evolution against exact continuous-time dynamics.

Using Krylov subspace exponentiation
(\texttt{scipy.sparse.linalg.expm\_multiply}) on the full
65,536-dimensional Hilbert space ($n_{\rm bdy}=8$, $\chi=4$),
we computed the late-time geometry correlation for continuous
Hamiltonian evolution. As shown in Table~\ref{tab:dt_convergence},
the exact continuous-time evolution yields an even stronger
geometric correlation ($r = +0.813$) than the $dt=0.1$ Trotter
approximation ($r = +0.778$).

These results confirm that the geometry protection is a robust
physical feature of the continuum XXZ Hamiltonian. The discrete
Trotter steps introduce artificial high-frequency heating that
weakly degrades the MBL protection, meaning the $dt=0.1$
parameter sweeps presented in the main text represent a strict,
conservative lower bound on the geometry-preserving capacity of
the MBL phase.

\begin{table}[h]
\centering
\caption{Convergence of late-time MI/lattice correlation
($W=30$, $\Delta=50$) evaluated over 5~seeds, comparing exact Krylov
exponentiation to second-order Trotter decomposition.}
\label{tab:dt_convergence}
\begin{tabular}{lcc}
\hline\hline
Method & Late-time $r$ & Deviation from Exact \\
\hline
\textbf{Exact (Krylov)} & $\bm{+0.813 \pm 0.004}$ & -- \\
Trotter $dt=0.02$ & $+0.803 \pm 0.003$ & $-0.011$ \\
Trotter $dt=0.05$ & $+0.786 \pm 0.007$ & $-0.027$ \\
Trotter $dt=0.10$ & $+0.778 \pm 0.006$ & $-0.035$ \\
Trotter $dt=0.20$ & $+0.734 \pm 0.012$ & $-0.079$ \\
\hline\hline
\end{tabular}
\end{table}

\bibliography{refs}

\end{document}